\documentclass[12pt]{article} 

\usepackage[utf8]{inputenc} 

\usepackage{geometry} 
\usepackage{graphicx} 
\RequirePackage[colorlinks,citecolor=blue,urlcolor=blue]{hyperref}
\RequirePackage[sort,round]{natbib}
\usepackage{amsmath, amssymb}
\usepackage{palatino}
\usepackage{float}

\usepackage{booktabs} 
\usepackage{array} 
\usepackage{paralist} 
\usepackage{verbatim} 
\usepackage{subfig} 
\usepackage{amsthm}
\newtheorem{theorem}{Theorem}

\newtheorem{example}{Example}
\newtheorem{remark}{Remark}

\usepackage{comment}
\usepackage{enumitem}

\usepackage{fancyhdr} 
\usepackage{sectsty}

\usepackage[nottoc,notlof,notlot]{tocbibind} 
\usepackage[titles,subfigure]{tocloft} 

\def\bSig\mathbf{\Sigma}

\def\PIT{\textsc{pit }}

\def\ARMA{\textsc{arma }}
\def\MA{\textsc{ma }}
\def\GLARMA{\textsc{glarma }}
\def\MELE{\textsc{mele }}
\def\MLE{\textsc{mle }}
\def\LRT{\textsc{lrt }}
\def\GLM{\textsc{glm }}
\def\GLMs{\textsc{glm}s }
\usepackage{amsfonts}

\newcommand{\R}{\mathbb{R}}

\begin{document}

\title{\Large{Semiparametric generalized linear models for time-series data}}
\author{\normalsize{{\bf Thomas Fung}$^a$ and {\bf Alan Huang}$^b$} \vspace{3mm} \\
\normalsize{$^a$Macquarie University and $^b$University of Queensland} \\
\normalsize{thomas.fung@mq.edu.au, alan.huang@uq.edu.au}}
\date{\normalsize{3 November, 2015}} 

\maketitle


\begin{abstract}

Time-series data in population health and epidemiology often involve non-Gaussian responses. In this note, we propose a semiparametric generalized linear models framework for time-series data that does not require specification of a working conditional response distribution for the data. Instead, the underlying response distribution is treated as an infinite-dimensional parameter which is estimated simultaneously with the usual finite-dimensional parameters via a maximum empirical likelihood approach. A general consistency result for the resulting estimators is given. Simulations suggest that both estimation and inferences using the proposed method can perform as well as correctly-specified parametric models even for moderate sample sizes, but can be more robust than parametric methods under model misspecification. The method is used to analyse the Polio dataset from \cite{Zeger1988} and a recent Kings Cross assault dataset from \cite{MWKF2015}. \vspace{2mm} 
\end{abstract}
%



%

\section{Introduction}
\label{intro}

\begin{figure}[H]
\begin{center}
\includegraphics[width=\textwidth, trim=3cm 1cm 1cm 1cm, clip=T]{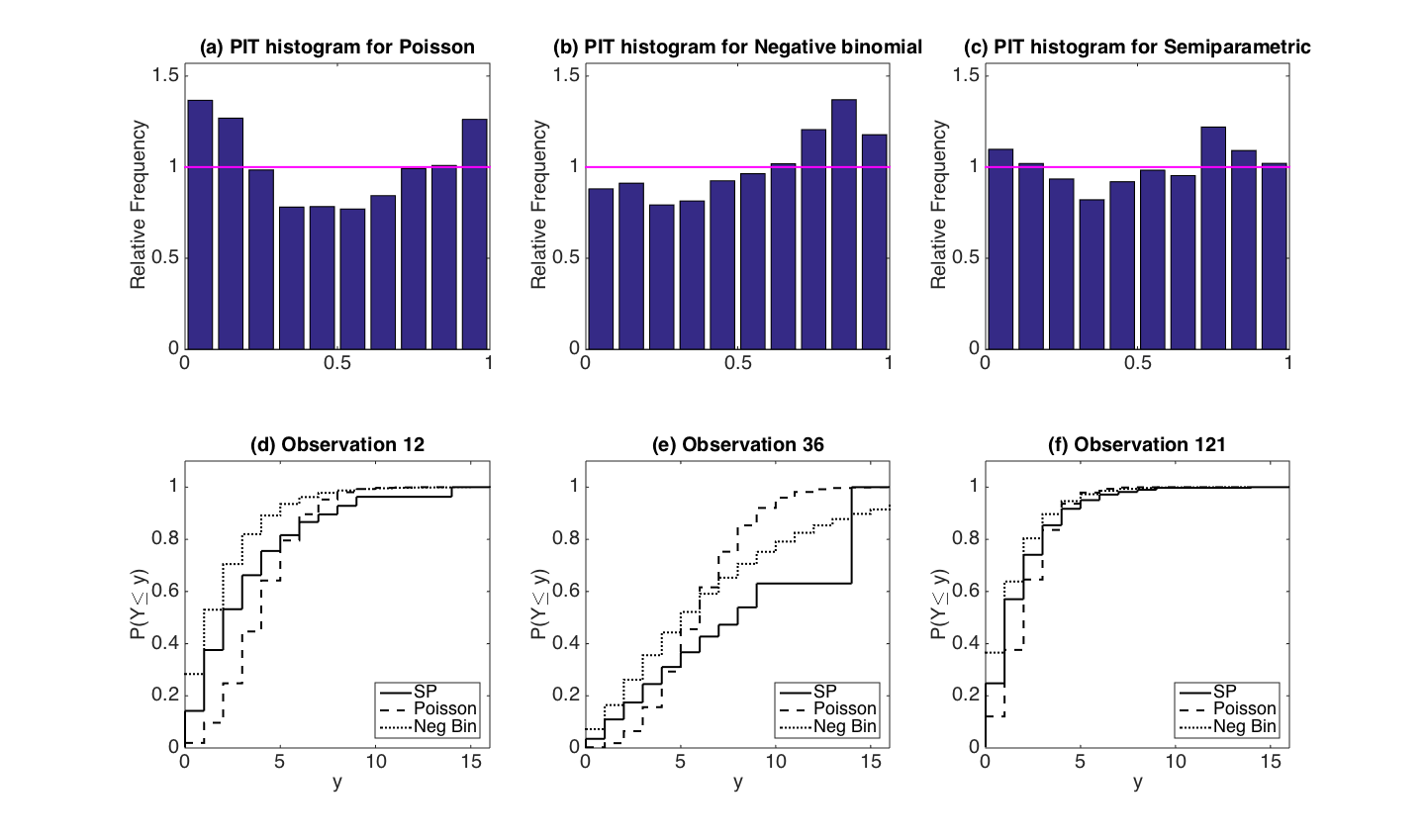}
\end{center}
\vspace{5mm}
\caption{Goodness-of-fit plots for the Polio dataset. Panels (a)--(c) are histograms of the probability integral transforms (\PIT) for Poisson, negative-binomial and semiparametric fits to the data. Panels (d)--(f) display the fitted conditional response distributions for the 12th, 36th and 121st observations using the Poisson, negative-binomial (Neg-Bin) and semiparametric (SP) models} 
\label{figure:polio pit}
\end{figure}

Datasets in population health and epidemiology often involve non-Gaussian responses, such as incidence counts and time-to-event measurements, collected over time. Being outside the familiar Gaussian autoregressive and moving average (\ARMA) framework makes modelling inherently difficult since it is not always apparent how to specify suitable conditional response distributions for the data. Consider, for example, the Polio dataset from \cite{Zeger1988}, consisting of monthly counts of poliomyelitis cases in the USA from year 1970 to 1983 as reported by the Centre of Disease Control. \cite{DDW2000} analysed this dataset using an observation-driven Poisson \ARMA model, but find evidence of overdispersion in the counts. To remedy this, \cite{DW2009} considered a negative-binomial model for the data. However, probability integral transformation plots, which should resemble the histogram of a standard uniform sample if the underlying distribution assumptions are appropriate, show noticeable departures from uniformity for both the Poisson and negative-binomial models (see panels (a) and (b) of Fig. \ref{figure:polio pit} and Section 5.1). This suggests that neither distribution fits the data adequately. A flexible yet parsimonious framework for modelling non-Gaussian time-series datasets is therefore warranted for such situations where specification of the response distribution is difficult.

This paper concerns the class of generalized linear models (\textsc{glm}s) for time-series data. More precisely, we consider models of the form 
\begin{equation}
\label{eq:model1}
Y_t | \mathcal{F}_{t-1} \sim \mbox{ExpFam}\{\mu(\mathcal{F}_{t-1}; \beta,\gamma)\} \ ,
\end{equation}
where $\mathcal{F}_{t-1}$ is the state of the system up to time $t$, which may include previous observations $Y_0,\ldots, Y_{t-1}$ and previous and current covariates $X_0, \ldots,X_{t}$, $\mu(\mathcal{F}_{t-1}; \beta, \gamma) = E(Y_t| \mathcal{F}_{t-1} )$ is a conditional mean function that may depend on a finite vector of parameters $(\beta\,^T,\gamma^T) \in \mathbb{R}^{q + r}$, and ExpFam$\{\mu\}$ is some exponential family indexed by its mean $\mu$. For clarity of exposition, we have split the mean-model parameter into two vectors here, with $\beta \in \mathbb{R}\,^q$ corresponding to the systematic component and $\gamma \in \mathbb{R}^r$ for the dependence structure of the model. We may also write $\mu_t \equiv \mu_t(\beta, \gamma)$ interchangeably for $\mu(\mathcal{F}_{t-1}; \beta, \gamma)$ whenever it is more  convenient to do so. 
Some specific models of the form (\ref{eq:model1}) have been discussed in \cite{KF2002}.

The framework (\ref{eq:model1}) is broad and encompasses, for example, Gaussian \ARMA models for continuous data \citep[e.g.,][Chapter 3]{BD2009}, Poisson models for count data \citep[e.g.,][]{FRT2009} and gamma and inverse-Gaussian models for time-to-event data \citep[e.g.,][]{AB1999, GJ2006}. The following are three explicit models covered by this framework, with the third and perhaps most interesting example examined in more detail throughout this paper.

\begin{example}[Gaussian \ARMA models]
\label{ex:1}
A process $\{Y_t\}$  is said to follow a Gaussian \textsc{arma}(1,1) process if it is stationary and satisfies
$$
Y_t = d+ \phi Y_{t-1}+Z_t+\psi Z_{t-1}
$$
where $\{Z_{t}\}$ is a gaussian white noise process. Here, the exponential family in (\ref{eq:model1}) is the normal and the mean-model parameters are $\beta=d$ for the systematic component and $\gamma^T = (\phi, \psi)$ for the autoregressive and moving average parameters.  
\end{example}

\begin{example}[Poisson autoregression]
\label{ex:2}
\citet{Fokianos2011} recently reviewed the autoregressive Poisson model $
Y_t | \mathcal{F}_{t-1} \sim \mbox{Poisson}\{\mu_t\}
$ for count time-series data, where
$
\mu_t = d +  a \mu_{t-1} + b Y_{t-1}
$
 for a linear mean-model, or 
$
\log(\mu_t) = d +  a \log(\mu_{t-1}) + b \log(Y_{t-1}+1)
$ 
for a log-linear mean-model. Here, the particular exponential family in (\ref{eq:model1}) is the Poisson and the mean-model parameters are $\beta = d$ for the systematic component and $\gamma^T = (a,b)$ for the autoregressive component.
\end{example}

\begin{example}[Observation-driven Poisson models]
\label{ex:3}
\cite{DDS2003} introduced the following observation-driven model for count data $Y_t$ with covariates $x_t$,
$$
Y_t | \mathcal{F}_{t-1} \sim \mbox{Poisson}\left\{\mu_t=e^{W_t}\right\}\,.
$$
It is further assumed that the state equation $W_t$ follows the model 
\begin{eqnarray*}
W_t &=& x_t^T \beta + Z_t, \quad Z_t  = \phi_1(Z_{t-1} + e_{t-1}) +\ldots +\phi_s(Z_{t-s} + e_{t-s}) + \psi_1 e_{t-1} +\ldots+\psi_q e_{t-q}, \\
e_t &=& (Y_t - e^{W_t})/\mbox{var}(Y_t|\mathcal{F}_{t-1})^\lambda = (Y_t - e^{W_t})/e^{\lambda W_t}\quad  \mbox{for some fixed } \lambda \in (0,1] \ .
\end{eqnarray*}
Here, the particular exponential family in (\ref{eq:model1}) is the Poisson, the mean-model parameters are $\beta$ for the systematic regression component and  $\gamma^T = (\phi^T,\psi^T)$ for the autoregressive and moving average component. This model is a special case of a generalized linear autoregressive and moving average (\textsc{glarma}) model. Note that the entire \GLARMA  framework is covered by (\ref{eq:model1}).
\end{example}

While the fully parametric models in Examples \ref{ex:1}--\ref{ex:3} have proven useful for analyzing particular datasets, the requirement of a particular exponential family for the conditional response distribution may be too restrictive for general problems. Indeed, one only has to look at, for example, the myriad of quasi-Poisson \citep[][Chapter 4]{KF2002}, zero-inflated Poisson \citep{YZC2013} and negative-binomial models \citep{DW2009} to see that the basic Poisson model is often inadequate for modeling count data. Note also that these alternative models generally do not form exponential families and are therefore outside framework (\ref{eq:model1}).

This paper introduces a parsimonious approach for modeling time-series data that is flexible and robust, yet remains completely within the exponential family framework (\ref{eq:model1}). The proposed approach is based on a novel exponential tilt representation of generalized linear models (\textsc{glm}s), introduced in \cite{RG2009}, that allows the underlying exponential family in (\ref{eq:model1}) to be unspecified. Instead, the particular distribution generating the exponential family is treated as an infinite-dimensional parameter, which is then estimated simultaneously from data along with the usual model parameters $(\beta,\gamma)$. 

The semiparametric framework we propose is rather generic, and can be used to relax distributional assumptions in essentially any parametric time-series model based on an exponential family. The usefulness of the proposed approach in practice is demonstrated by fitting it to the polio dataset from \cite{Zeger1988} as well as a more modern application to the Kings Cross assault dataset from \cite{MWKF2015}.

\section{Semiparametric GLMs for time-series data}
Using the exponential tilt representation of \citet{RG2009}, we can write model (\ref{eq:model1}) as
\begin{equation}
Y_t | \mathcal{F}_{t-1} \sim \exp(b_t + \theta_t y) dF(y) \ , \quad t=\ldots, 0,1,2,\ldots \ ,
\label{eq:model}
\end{equation}
where $F$ is some response distribution, $\{b_t\}$ are normalization constants satisfying
\begin{equation}
b_t = -\log \int \exp(\theta_t y) dF(y) \ , \quad t=\ldots, 0, 1,2 \ldots \ ,
\label{eq:normconstr}
\end{equation}
and the tilts $\{\theta_t\}$ are determined by the mean-model via
$$
\mu_t(\beta,\gamma) = E(Y_t|\mathcal{F}_{t-1}) = \int y \exp(b_t + \theta_t y) dF(y) \ , \quad t=\ldots, 0, 1, 2,\ldots .
$$
This last equation can also be rearranged as
\begin{equation}
0 = \int [y-\mu_t(\beta,\gamma)] \exp(\theta_t y) dF(y) \ , \quad t=\ldots, 0,1,2 \ldots,
\label{eq:meanconstr}
\end{equation}
which avoids explicit dependence on $b_t$. 

\begin{remark} Note that $F$ is not the stationary distribution of the process. Rather, it is the underlying distribution generating the exponential family.
\end{remark}

Classical time-series models, such as those based on normal, Poisson or gamma distributions, can be recovered by setting $F$ to be a normal, Poisson or gamma distribution, respectively. The main advantage of the exponential tilt approach, however, is that the underlying distribution $F$ can be left 
unspecified and instead treated as an infinite-dimensional parameter along with the usual mean-model parameters $(\beta,\gamma)$ in the log-likelihood function,
$$
l(\beta,\gamma, F) = \sum_{t=0}^n \left\{ \log dF(Y_t) + b_t + \theta_t Y_t \right\}
$$
with $b_t$ given by (\ref{eq:normconstr}) and $\theta_t$ given by (\ref{eq:meanconstr}), for  $t=0,1,\ldots, n$. 

Treating $F$ as a free parameter gives us much flexibility in model specification. For example, overdispersed or underdispersed counts can be dealt with simply by $F$ having heavier or lighter tails than a nominal Poisson distribution. Similarly, zero-inflated data can be dealt with by $F$ having an additional probability mass at zero. Two examples of non-standard exponential families covered by our framework are displayed in Fig. 1 in the Online Supplement. These include a zero-inflated Poisson family and a normal-mixture family.


It is important to note that our framework is semiparametric as it contains infinitely many exponential families, one for each unique underlying distribution $F$. We are by no means restricted to the handful of classical families, such as the normal, Poisson and gamma families. Indeed, instead of specifying $F$ {\it a priori},
the key innovation in this paper is to allow the underlying distribution $F$ to be estimated simultaneously along with $\beta$, thereby letting the data inform us about which distribution fits best. While there are conceivably many methods that can be used to carry out this joint estimation, this paper proposes a maximum empirical likelihood approach. This is described in more detail in Section 3. Some other theoretical properties of the model are given in the Appendix.


\section{Estimation via maximum empirical likelihood}
\subsection{Methodology}
To fit model (\ref{eq:model})--(\ref{eq:meanconstr}) to data, consider replacing the unknown $dF$ with a probability mass function $\{p_0,\ldots, p_n\}$ on the observed support $Y_0, \ldots, Y_n$ with $\sum_{j=0}^n p_j = 1$, so that the log-likelihood becomes the following empirical log-likelihood:
\begin{eqnarray}
\label{eq:emplik}
l(\beta,\gamma, p) &=&  \sum_{t=0}^n \left\{ \log p_t + b_t + \theta_t Y_t \right\} \ , \\
\label{eq:empnorm}
\mbox{with } b_t &=& -\log \sum_{j=0}^n \exp(\theta_t Y_j) p_j \ , \hspace{19mm} t=0, 1,\ldots, n \ , \\
\label{eq:empmean}
\mbox{and } 0 &=& \sum_{j=0}^n [Y_j - \mu_t( \beta,\gamma)] \exp(\theta_t Y_j) p_j \ , \hspace{6mm} t=0,1,\ldots, n\ .
\end{eqnarray}
Note that equations (\ref{eq:empnorm}) and (\ref{eq:empmean}) are the empirical versions of the normalization and mean constraints (\ref{eq:normconstr}) and (\ref{eq:meanconstr}), respectively.  A maximum empirical likelihood estimator (\textsc{mele}) for $(\beta,\gamma, p)$ can then be defined as
$$
(\hat \beta, \hat \gamma, \hat p) = \mbox{argmax}_{\beta,\gamma, p} l(\beta,\gamma, p),
$$ 
subject to constraints (\ref{eq:empnorm}) and (\ref{eq:empmean}) and the corresponding \MELE for $F$ is given by the cumulative sum
$$
\hat F(y) = \sum_{j=0}^n \hat p_j 1(Y_j \le y)  \ .
$$
This also means that the \MELE for the cdf of a particular response $Y_t$ is 
$$
\hat{F}_{Y_t}(y) = \sum^n_{j=0} \hat{p}_j \exp(b_t+\theta_t Y_j) 1(Y_j \leq y).
$$

Clearly the choice of regressors will have a big impact on the behaviour of $\mu_t(\beta,\gamma)$ and subsequently the properties of the estimator. \cite{DDW2000} and \cite{DW2009} discuss a list of conditions such that a regressor can be incorporated. The conditions are fairly general and hold for a wide range of covariates, including a trend function $x_{nt} = f(t/n)$ where $f(\cdot)$ is a continuous function defined on the unit interval $[0,1]$ and harmonics functions that specify annual, seasonal or day-of-the-week effects.

The following theorem establishes a general consistency result for the proposed estimator. In particular, it states that whenever consistency holds for the correctly-specified parametric maximum likelihood estimator (\textsc{mle}), 
then it also holds for the semiparametric \textsc{mele}.
The proof of Theorem \ref{thm:consistency} is provided in the Supplementary Materials. 

\begin{theorem}[Consistency]
{\color{black} Assume that the joint process $(Y_t, W_t)$ is stationary and uniformly ergodic, and that conditions C1--C3 in Appendix \ref{sec:regcond} hold.}
Then, whenever the (correctly-specified) parametric \MLE of $(\beta,\gamma)$ is consistent, so is the semiparametric \MELE $(\hat \beta, \hat \gamma)$. Moreover, 
$\sup_{y \in \mathcal{Y}} |\hat F(y) - F^*(y)| \to 0$ in probability as $n \to \infty$, where $F^*$ is the true underlying distribution.
\label{thm:consistency}
\end{theorem}

Establishing stationarity, ergodicity and consistency even in relatively simple parametric settings is still ongoing research \citep[e.g.][]{DDW2000}. Theorem \ref{thm:consistency} implies that whenever advances are made about the consistency of particular parametric estimators, then such results automatically apply to the semiparametric estimator also.

%

Note that our 
maximum empirical likelihood estimation scheme is generic for any mean-model $\mu_t$. Specific models for $\mu_t$ may require additional constraints and calculations, the complexity of which depends on the complexity of the systematic and dependency structures of the mean-model. To illustrate this, we make explicit below the optimization problem for computing the semiparametric \MELE for the \GLARMA model from Example \ref{ex:3}. A {\tt MATLAB} program carrying out these calculations can be found in {\tt spglarma3.m}, which is obtainable by emailing the authors. An alternative version, {\tt spglarma3par.m}, utilizes the parallel computing toolbox for faster computations. 
The corresponding calculations for the fully parametric Poisson model can be found in \citet[][Section 3.1]{DDS2003}.

\vspace{3mm}
\noindent E\textsc{xample} 1.3 (continued). The constrained optimization problem for obtaining the \MELE of $(\beta, \phi, \psi, F)$ in the \GLARMA model of \citet{DDS2003} with $\lambda = 1/2$ is:
\vspace{-2mm} {
\begin{eqnarray*}
\hline
\mbox{maximize } l(\beta, \phi, && \hspace{-4mm} \psi, p; b, \theta) =  \sum_{t=0}^n \left\{ \log p_t + b_t + \theta_t Y_t \right\} \mbox{ over } \beta, \phi, \psi, p, b\mbox{ and } \theta \ , \\
\mbox{subject to } \hspace{5mm} 0 &=& \sum_{j=0}^n \left[Y_j - \exp{( x_t^T\beta + Z_t)}\right] \exp(\theta_t Y_j) p_j \ , \  t=0, \ldots, n, \hspace{0mm} \mbox{ (mean constraints)}\\
b_t &=& -\log \sum_{j=0}^n \exp(\theta_t Y_j) p_j \ , \qquad t=0,\ldots, n, \hspace{8mm} \mbox{ (normalization constraints)}\\
\mbox{where \hspace{1cm}} 
Z_t &=& \phi_1(Z_{t-1}+e_{t-1}) + \cdots+ \phi_s(Z_{t-s}+e_{t-s}) \\
&& \hspace{34mm}+\psi_1e_{t-1}+\cdots+\psi_qe_{t-q} \ , \hspace{13mm} \mbox{ (\ARMA constraints)} \\
e_t &=& \frac{Y_t- \exp{( x_t^T\beta + Z_t)}}{\sqrt{\mbox{var}(Y_t|\mathcal{F}_{t-1})}}, \hspace{54mm} \mbox{ (Pearson residuals)} \\
\mbox{var}(Y_t|\mathcal{F}_{t-1}) &=& \sum_{j=0}^n \left[Y_j - \exp{( x_t^T\beta + Z_t)}\right]^2 \exp{(b_t + \theta_t Y_j)}p_j \ . \hspace{13mm} \mbox{ (variance function)} \\
\mbox{The \MELE for $F$ is } && \hspace{-5mm} \mbox{then obtained by the cumulative sum }
\hat F(y) = \sum_{t=0}^n \hat p_t 1(Y_t \le y) .\\
\hline
\end{eqnarray*}}
\noindent An alternative to Pearson residuals is score-type residuals, obtained by setting $\lambda=1$: 
$$
\hspace{3cm} e_t = \frac{Y_t- \exp{( x_t^T\beta + Z_t)}}{\mbox{var}(Y_t|\mathcal{F}_{t-1})}. \hspace{57mm} \mbox{(Score-type residuals)}
$$
The method of \cite{CKL2008} can also be used.

\subsection{Simulation Studies}
\label{se:sim1}
To examine the practical performance of the proposed semiparametric \MELE under both correct specification and misspecification, we simulate from different \GLARMA models based on Example \ref{ex:3} and compare the results with the \MLE from parametric models. 

The first set of simulations is similar to those in \cite{DDS2003} which is based on a Poisson distribution with the log-link,
$$
Y_t| \mathcal{F}_{t-1} \sim \mbox{Poisson}\{\mu_t = e^{W_t}\},
$$
under two settings, namely, one corresponding to a stationary trend,
$$ \mbox{Model 1:} \quad W_t = \beta_0 + \psi(Y_{t-1}-e^{W_{t-1}})e^{-\frac{1}{2}W_{t-1}} \ ,$$
and the other with a linear trend,
$$ \mbox{Model 2:} \quad W_t = \beta_0 +\beta_1t/n+ \psi(Y_{t-1}-e^{W_{t-1}})e^{-\frac{1}{2}W_{t-1}}.$$
The sample sizes were $n=250$ with $N =1000$ replications for both settings. A burn-in period of length 100 was used in the simulation of the realisations. For each simulated dataset, we fit both the correctly-specified Poisson model, via the {\tt glarma} package in {\tt R} of \citet{DLS2014} with the default options (Pearson residuals and Fisher Scoring method), and the semiparametric model, via the {\tt spglarma3par} function in {\tt MATLAB}. 

The densities of the parameter estimates across the simulations for both methods are plotted in Fig.~2 in the Online Supplement. As both models are correctly specified here, it is not surprising to see they exhibit almost identical performance for estimating all model parameters.

It is perhaps more interesting to look at the performance of the estimators under model misspecification. To this end, a second set of simulations was carried out based on a negative-binomial distribution with the log-link,
$$ 
Y_t|\mathcal{F}_{t-1} \sim \mbox{NegBin}(\mu_t = e^{W_t}, \alpha),$$
such that the conditional density function of $Y_t$ is 
$$
P(Y_t = y_t | \mathcal{F}_{t-1}) = \frac{\Gamma(y_t+\alpha)}{\Gamma(\alpha)y_t!}\left(\frac{\alpha}{\alpha+\mu_t}\right)^{\alpha}\left(\frac{\mu_t}{\alpha+\mu_t}\right)^{y_t}, \quad \mbox{for $y_t = 1, \ldots$},$$
where 
\begin{eqnarray*}
\mbox{Model 3:} \quad W_t &=& \beta_0 + \beta_1t/n + \beta_2\cos(2\pi t/6)+\beta_3 \sin(2\pi t/6) + Z_t,\\
Z_t &=& \phi( Z_{t-1} + e_{t-1}), \quad e_{t-1} = \frac{Y_{t-1} - e^{W_{t-1}}}{\sqrt{e^{W_{t-1}}+e^{2W_{t-1}}/\alpha}}.
\end{eqnarray*}
The linear predictor consists of a linear trend, two harmonic functions as well as an $\textsc{ar}(1)$ component. The sample size is $n=500$ with $N=1000$ replications. The parameter values are set as $(\beta_0, \beta_1, \beta_2, \beta_3, \phi,\alpha) = $(0$\cdot$1, 0$\cdot$2, 0$\cdot$3, 0$\cdot$4, 0$\cdot$25, 4). A burn-in period of length 100 was used in the simulation of the realisations. Note that the negative-binomial distribution does not correspond to a \GLM unless the dispersion parameter $\alpha$ is known {\it a priori}.
 
\begin{figure}[h!]
\begin{center}
\begin{tabular}{ccc}
\scalebox{0.25}{\includegraphics[trim = 1cm 0 1cm 0, clip=true]{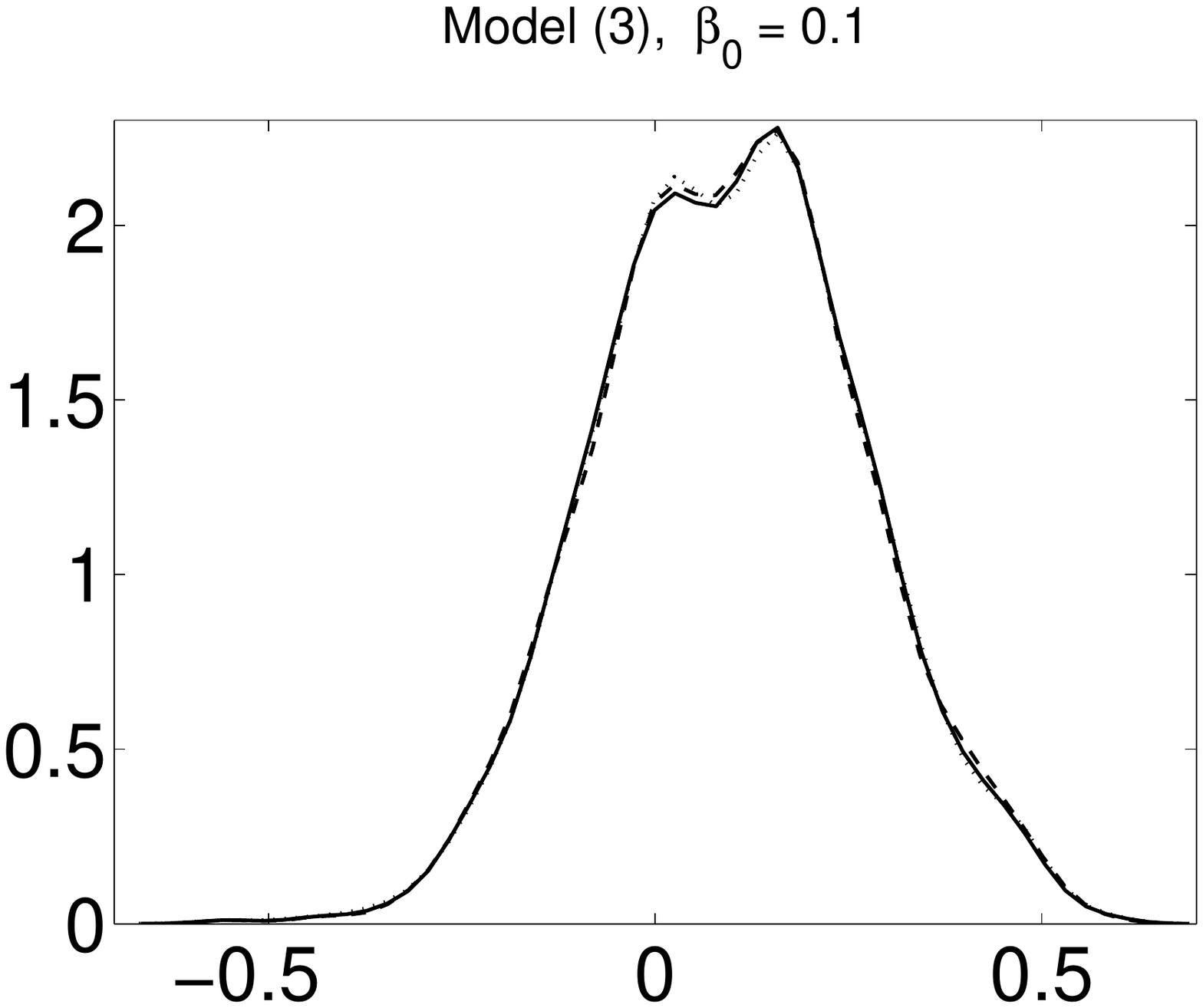}} & \scalebox{0.25}{\includegraphics[trim = 1cm 0 1cm 0, clip=true]{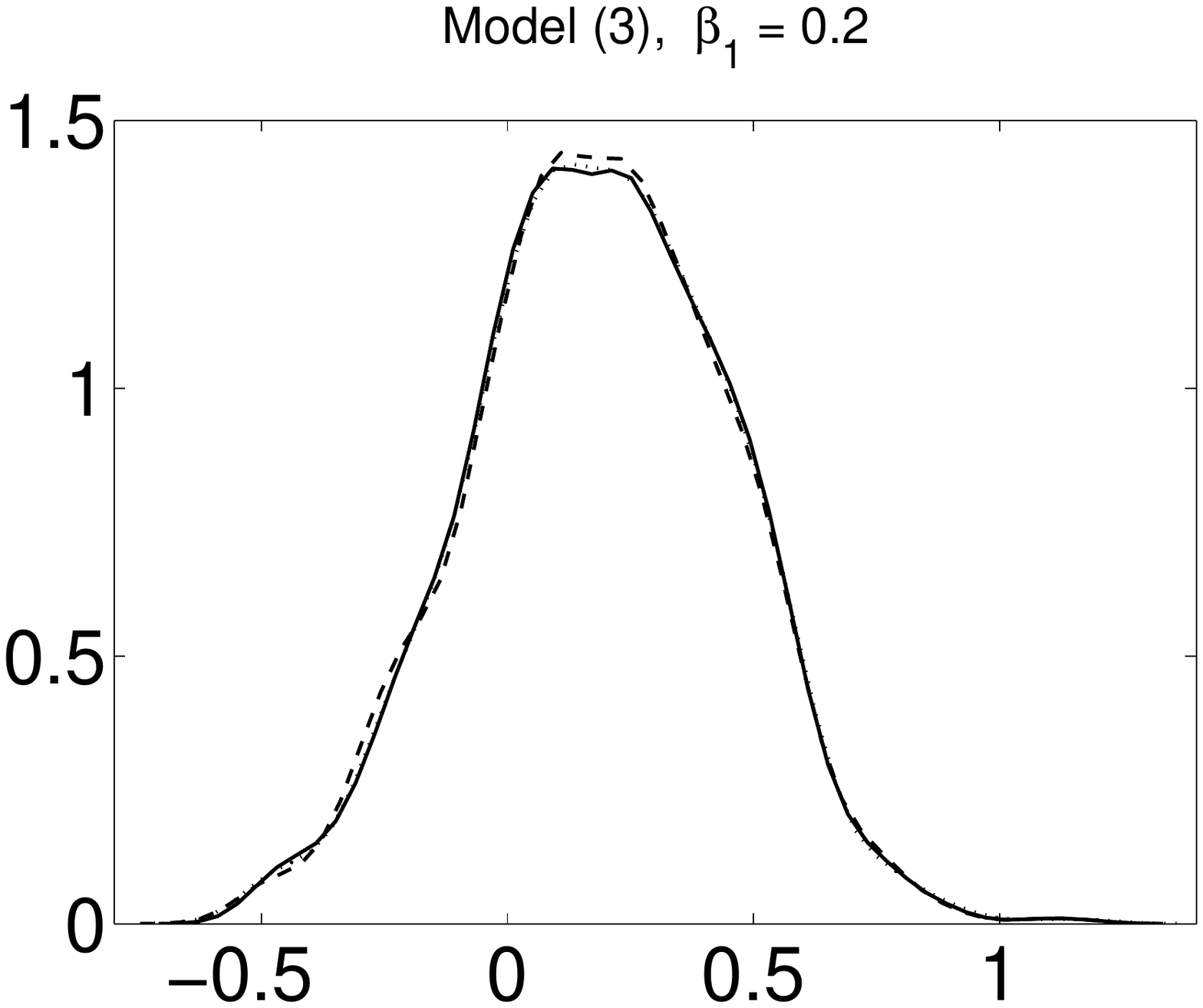}}\\[-3cm] 
\scalebox{0.25}{\includegraphics[trim = 1cm 0 1cm 0, clip=true]{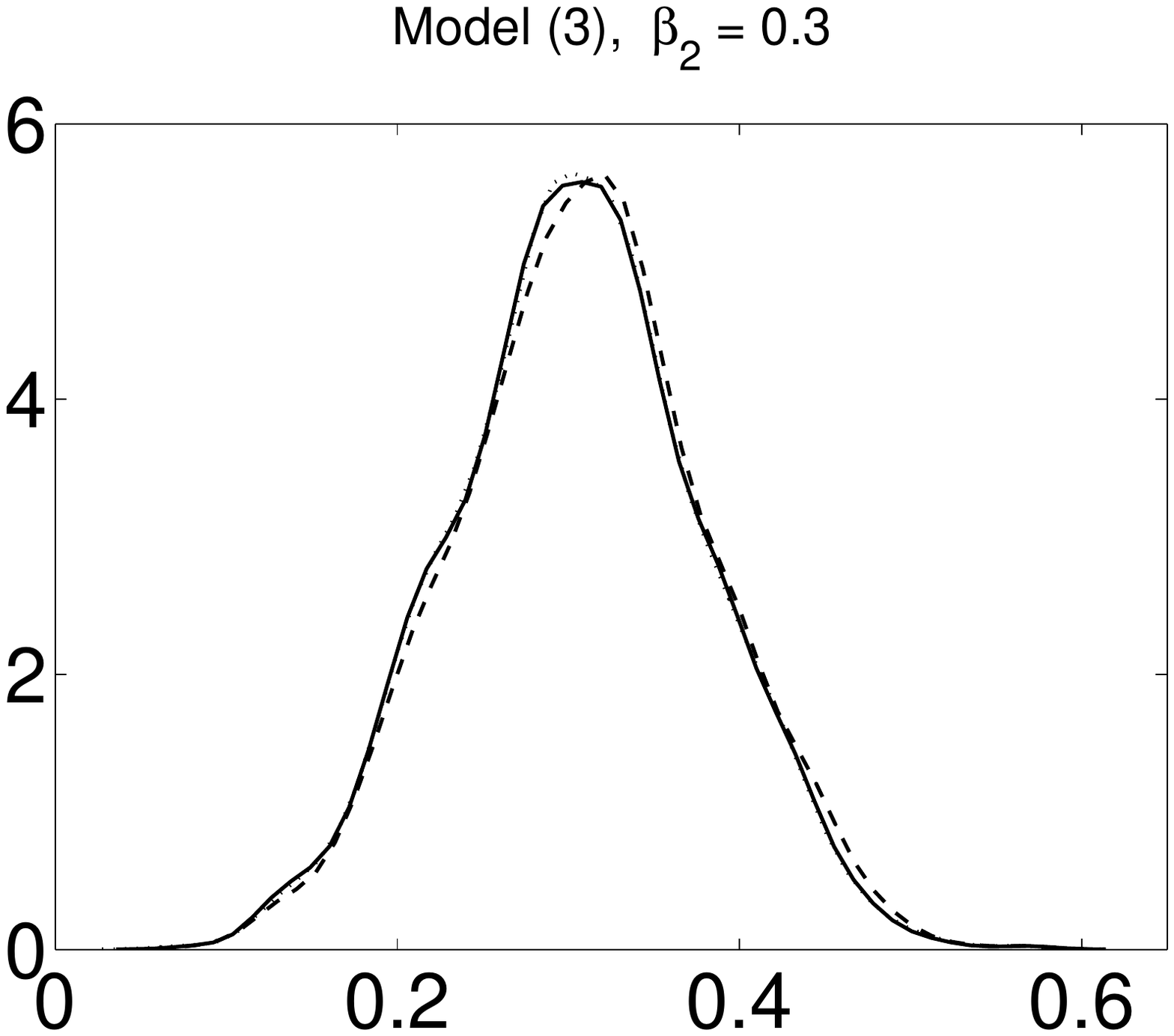}} & \scalebox{0.25}{\includegraphics[trim = 1cm 0 1cm 0, clip=true]{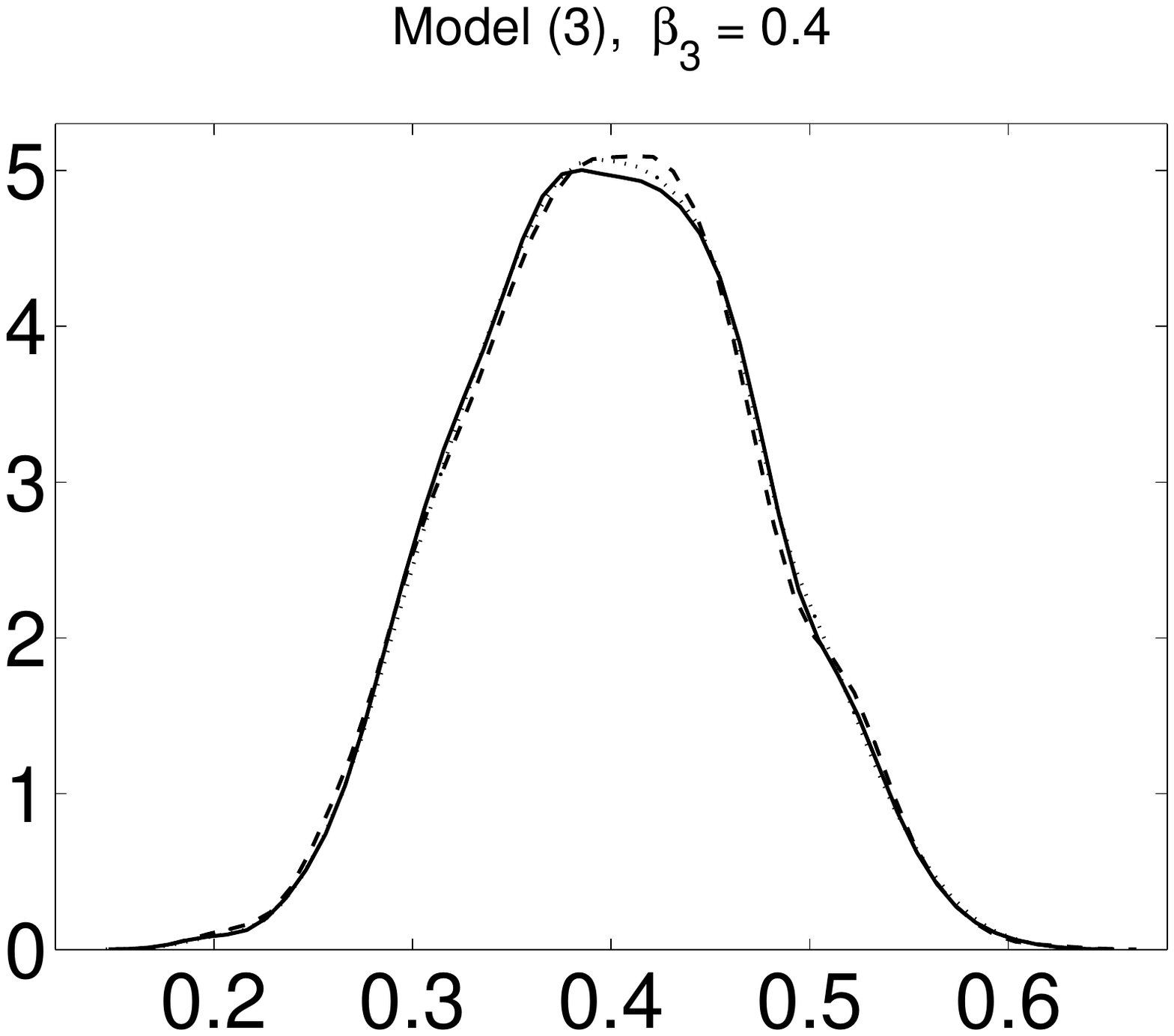}} & \scalebox{0.25}{\includegraphics[trim = 1cm 0 1cm 0, clip=true]{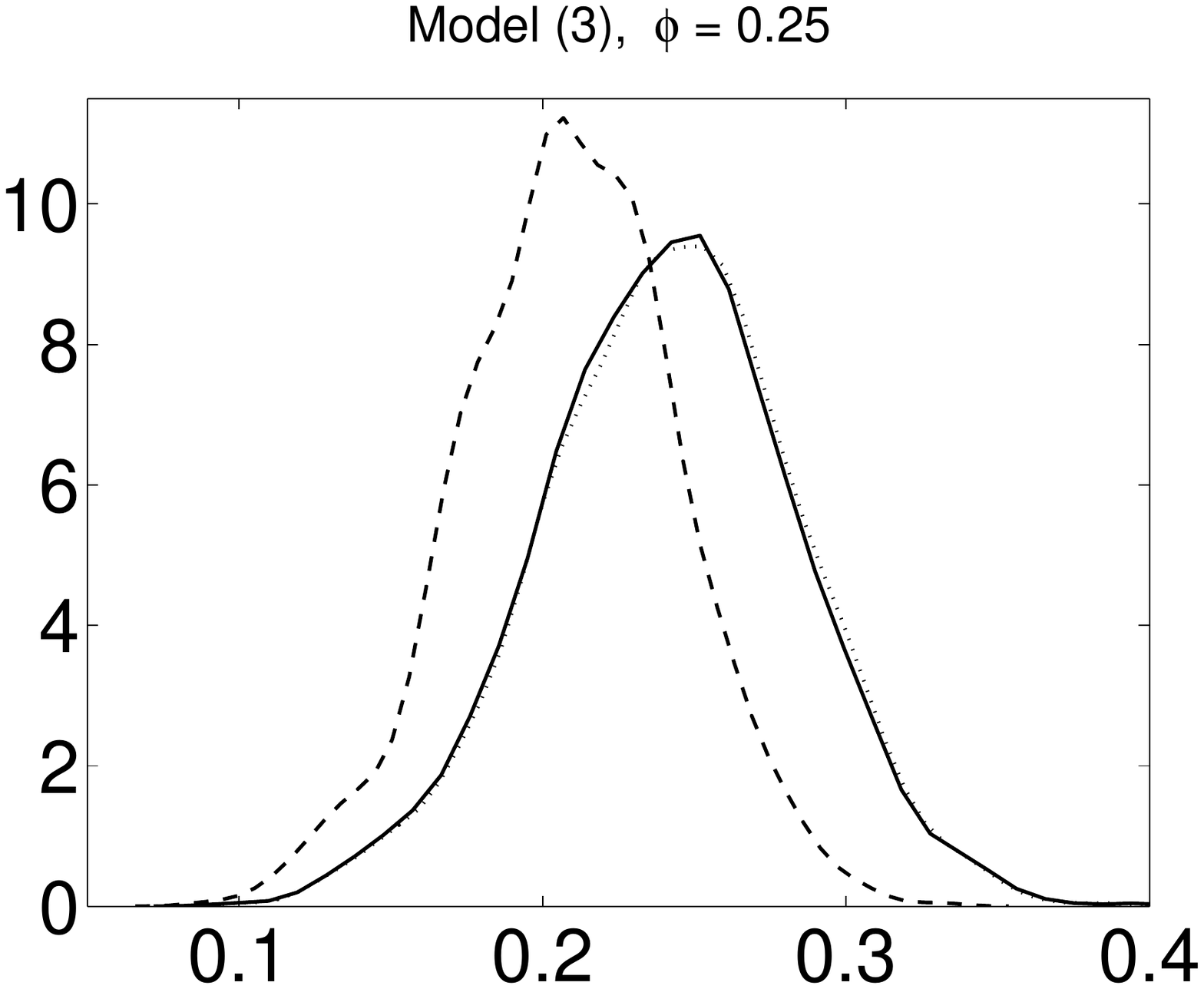}}
\end{tabular}
\caption{Empirical densities of the maximum likelihood estimators for Model (3) using the semiparametric (solid), misspecified Poisson (dashed) and correctly-specified negative-binomial (dotted) models.} \label{Figure: Neg bin densities}
\end{center}
\end{figure}

For each simulated dataset, we fit the correctly-specified negative-binomial model and the misspecified Poisson model using the {\tt glarma} package in {\tt R}. We also fit the proposed semiparametric model. The densities of the parameter estimates across the simulations for all three methods are plotted in Fig. \ref{Figure: Neg bin densities}. Although the misspecified Poisson model can give good estimates for the regression coefficients $(\beta_0,\beta_1, \beta_2, \beta_3)$, it gives very different and severely biased estimates for the autoregression coefficient $\phi$ (see subpanel 5). Note that the autoregressive component of the model here is intimately connected with the overall error structure, so it is perhaps not surprising that the misspecification of the conditional error distribution has led to particularly poor estimates of the autoregression parameter. In contrast, the proposed semiparametric method gives almost identical estimates as the correctly-specified negative-binomial model for both the regression and autoregressive coefficients, demonstrating the robustness and flexibility of the proposed method under model misspecification. 

{\color{black} The semiparameteric approach is computationally more demanding than parametric approaches, for obvious reasons. Models 2 ($n=250$) and 3 ($n=500$) took, on average, 19 and 67 seconds (Macbook Pro, 2.3 GHz Intel Core i7, 16 GB 1600 RAM) to fit, respectively, whereas the {\tt glarma} package took no more than a few seconds. 
The current software uses only the built-in \texttt{MATLAB} optimizer \texttt{fmincon}, and we expect an improvement in computation times as we implement our own algorithm.}

\section{Inference via the likelihood ratio test}
\subsection{Methodology}
For inferences on the model parameters $(\beta, \gamma)$, \citet[][Section 3.1]{DDS2003} suggested inverting the numerical Hessian of the log-likelihood function at the \MLE to estimate its standard errors. Here, we propose a more direct method based on the likelihood ratio test. 

Suppose we are interested in testing $H_0: \beta_k = 0$ for some component of $\beta$. To do this, we can fit the semiparametric \GLM with and without the constraint, obtaining the empirical log-likelihood ratio test (\textsc{lrt}) statistic,
$$
LRT = 2\left\{l(\hat \beta, \hat \gamma, \hat p) - \sup_{\beta_k = 0} l(\beta, \gamma, p) \right\} \ .
$$
From this, we can compute an ``equivalent standard error", denoted by $se_{eq}$, via
$$
LRT= \left(\frac{\hat \beta_k - 0}{se_{eq}}\right)^2 
$$
and then construct $100(1-\alpha)\%$ Wald-type confidence intervals via the usual $\hat \beta_k \pm z_{\alpha/2} se_{eq}$.  Note that the equivalent standard error is such that the corresponding $z$-test for testing $\beta_k = 0$ achieves the same significance level as the likelihood ratio test with a $\chi^2_1$ calibration. An identical procedure can be applied for inferences on the components of the \ARMA parameter $\gamma$.

In some scenarios, perhaps motivated by previous experience and/or theoretical arguments, a null value other than 0 can be used to calculate an equivalent standard error, provided it does not lie on the boundary of the parameter space. In the absence of any additional knowledge, however, the choice of 0 seems a good default to use in practice, as supported by the simulation results in Section 4.2. 


\subsection{Simulation Studies}
To check the accuracy of the likelihood ratio test as the basis for inferences, the following two expanded models were fitted to the data generated by Model 1 in Section 3.2.1: 
$$ \mbox{Model 1$'$:} \quad W_t = \beta_0 +\beta_1t/n+ \psi(Y_{t-1}-e^{W_{t-1}})e^{-\frac{1}{2}W_{t-1}},$$
and 
$$  \mbox{Model 1$''$:} \quad W_t = \beta_0 + \psi(Y_{t-1}-e^{W_{t-1}})e^{-\frac{1}{2}W_{t-1}} +\psi_2(Y_{t-2} -e^{W_{t-2}})e^{-\frac{1}{2}W_{t-2}}.$$
Note that the linear term $\beta_1 t/n$ in Model 1$'$ and the second order \textsc{ma} component $\psi_2(Y_{t-2} -e^{W_{t-2}})e^{-W_{t-2}/2}$ in Model 1$''$ are redundant. This is equivalent to testing $\beta_1=0$ and $\psi_2 = 0$, respectively. 

If the proposed inferences based on the likelihood ratio test are to be useful, then the \LRT statistic for both tests should have approximate $\chi^2_1$ distributions. The Type I error rates for testing $\beta_1 =0$  using the \LRT at nominal 10\%, 5\% and 1\%  levels were 12.3\%, 6.6\% and 1.6\%, respectively, across 1000 simulations. The corresponding Type I errors for testing $\psi_2 = 0$ were 10.0\%, 6.0\% and 1.2\%, respectively.
These results indicate that the asymptotic $\chi^2$ calibration is quite accurate even for moderate sample sizes.



\begin{table}
\begin{center}
\def~{\hphantom{0}}
\caption{Empirical and estimated equivalent standard errors from negative-binomial data (Model 3) using the semiparametric, Poisson and negative-binomial models. $se_{emp}=$ empirical standard deviation ($\times 10\,^2$), $\overline{se}_{eq} =$ average equivalent standard error ($\times 10\,^2$), $\overline{se}=$ average standard error ($\times 10\,^2$), and coverage = coverage rate (\%) of the 95\% confidence interval, across 1,000 simulations}{
\begin{tabular}{crrcrrcrrc}
& \multicolumn{3}{c}{semiparametric} & \multicolumn{3}{c}{Poisson}& \multicolumn{3}{c}{negative-binomial} \\
parameter &  $se_{emp}$ & $\overline{se}_{eq}$ & coverage& $se_{emp}$ & $\overline{se}\ $ & coverage & $se_{emp}$ & $\overline{se}\ $ & coverage\\  
$\beta_0$& 16.8 & 16.1   & 93.9& 16.9 & 13.7 & 89.0& 16.9 & 16.3 & 94.1 \\
$\beta_1$  & 26.1 & 25.0 & 93.8 & 26.2 & 21.3 & 88.9 & 26.3 & 25.5 & 94.3 \\
$\beta_2$ & 7.3 & 7.2 & 95.1 & 7.2& 7.3 & 95.4& 7.2& 7.3 & 95.4\\
$\beta_3$ & 7.1 & 7.2 & 96.5& 7.1& 7.3 & 96.6 & 7.1 & 7.2 & 96.5 \\
$\phi $ & 4.2 & 4.1 & 93.9  & 3.6& 2.8 & 66.3 & 4.2 & 4.2&  94.3
\end{tabular}}
\label{Table: model 3 equiv se compare}
\end{center}
\end{table}

Next, we examine the performance of the proposed equivalent standard error approach for constructing confidence intervals for the parameters in the negative-binomial time-series generated by Model 3. The results from our simulation are reported in Table \ref{Table: model 3 equiv se compare}. We see that the average equivalent standard errors are very close to their corresponding empirical standard deviations across the 1,000 parameter estimates. Table \ref{Table: model 3 equiv se compare} also displays the coverage rates of 95\% Wald-type confidence intervals using the equivalent standard error, with the results suggesting that the proposed method performs as well as the correctly-specified negative-binomial model. In contrast, the Poisson results are particularly poor and indicate that model misspecification can lead to severe biases for inferences on model parameters. This reiterates the value of flexible and robust modeling of the conditional error distribution.  We can conclude that the semiparametric equivalent standard errors can be used in much the same way as (correctly-specified) parametric standard errors for inferences on model parameters. 


\section{Data analysis examples}
\subsection{Polio dataset}
To demonstrate the proposed method, we reanalyze the polio dataset studied by \cite{Zeger1988}, which consists of monthly counts of poliomyelitis cases in the USA from year 1970 to 1983 as reported by the Centres for Disease Control. \citet{DDW2000} and \citet{DW2009} modelled the counts using the Poisson and negative-binomial distributions, respectively. Here, we specify the same set of covariates as in these earlier studies, with
$$
x_{t} = \{1,\, t'/1000,\, \cos(2\pi t'/12), \, \sin(2\pi t'/12),\, \cos(2\pi t'/6),\, \sin(2\pi t'/6)\}^T,$$
where $t' = t-73$ is used to center the intercept term at January 1976. We  specify a log-link and a \textsc{ma} process with lags 1, 2 and 5, following the \GLARMA approach of \citet{DDW2000}.

\begin{table}
\begin{center}
\def~{\hphantom{0}}
\caption{Polio dataset: Estimated coefficients, equivalent standard errors ($se_{eq}$) and standard errors ($se$) using the semiparametric, Poisson and negative-binomial models}{
\begin{tabular}{lrrrrrrr}
& \multicolumn{2}{c}{semiparametric} & \multicolumn{2}{c}{Poisson} & \multicolumn{2}{c}{negative-binomial} \\
Variable  &  estimate & $se_{eq}$ & estimate & $se$  & estimate & $se$ \\
Intercept & 0.149 & 0.137 & 0.130 & 0.112 & 0.180 & 0.144\\
Trend & -3.960& 2.771& -3.928 & 2.145 & -3.171 & 2.715\\
$\cos(2\pi t'/12)$ &-0.093 & 0.166 & -0.099 & 0.118 & -0.139 & 0.178\\
$\sin(2\pi t'/12)$ & -0.518& 0.188 & -0.531 & 0.138  & -0.544 & 0.205\\
$\cos(2\pi t'/6)$ & 0.281 & 0.147 & 0.211 & 0.111 & 0.353 & 0.150\\
$\sin(2\pi t'/6)$ & -0.277& 0.155& -0.393 & 0.116 & -0.380 & 0.153\\
\MA\unskip, lag 1 & 0.320& 0.109& 0.218 & 0.047 & 0.304 & 0.065\\
\MA\unskip, lag 2 &  0.221 & 0.098 & 0.127 & 0.047 & 0.238 & 0.080\\
\MA\unskip, lag 5 & -0.016 & 0.099& 0.087 & 0.042 & -0.068 & 0.076
\end{tabular}}
\label{Table: Polio}
\end{center}
\end{table}

The coefficient estimates from the Poisson, negative-binomial and semiparametric fit to the data, along with their estimated standard errors, are reported in Table \ref{Table: Polio}. Whilst the fitted coefficients are quite similar for all three approaches, the estimated standard errors from the Poisson model are significantly smaller than the other two. The larger standard errors from the negative-binomial and semiparametric fits reflect the overdispersion in the counts, as noted in \citet{DDW2000}.

The estimated conditional response distributions for the 12th, 36th and 121st observations, covering a range of conditional mean values, are displayed in panels (d)--(f) of Fig. \ref{figure:polio pit}. These plots indicate that the semiparametric fit to the data is distinctively different to both the Poisson and negative-binomial fits. 
We can also use histograms of the probability integral transformation (\PIT\unskip) to compare and assess the underlying distributional assumptions. The {\tt glarma} package uses the formulation in \cite{CGH2009} so that \PIT can be applied to a discrete distribution. If the distributional assumptions are correct, the histogram should resemble the density function of a standard uniform distribution. As the estimated response distribution is discrete under our semiparametric model, the implementation of \PIT for our approach is also straightforward. From the \PIT\unskip s of all three fitted models, plotted in panels (a)--(c) of Fig. \ref{figure:polio pit}, we see that the plot for the semiparametric model is closest to uniform, with the negative-binomial and Poisson models being noticeably inferior. This suggest that neither parametric model is adequate here. Subsequently, standard errors and inferences based on these parametric models may well be biased and unreliable.

%

\subsection{Kings Cross assault counts}
\cite{MWKF2015} recently studied the number of assaults in the Kings Cross area of Sydney. A major practical question for which the data were originally collected was to determine the effect of the January 2014 reforms to the NSW Liquor Act, designed to curb alcohol-related violence, to the number of incidence of assault in Kings Cross, NSW, Australia. The data consists of monthly counts of (police recorded non-domestic) assaults in the area from January 2009 to September 2014. A state space Poisson model for count data of \cite{DK2012} was used in the original study but 
we shall reexamine the data within the \GLARMA framework. 

We follow the original study by specifying a log-link and the same set of covariates which includes an intercept, a step intervention term ( from January 2014 onwards), a cosine harmonic term at a period of 12 months.  Specifically, 
$$
x_t = \{1,\,I(t\geq 60),\,\cos(2\pi t/12)\},\quad t=0,1,\ldots 68 ,
$$
where $I(.)$ is an indicator function. A \MA process with lags 1 and 2 was also included in the model. 

\begin{table}
\begin{center}
\def~{\hphantom{0}}
\caption{Kings Cross assault counts: Estimated coefficients, equivalent standard errors ($se_{eq}$) and standard errors ($se$) using the semiparametric and Poisson models}{
\begin{tabular}{lrrrrr}
 & \multicolumn{2}{c}{semiparametric} & \multicolumn{2}{c}{Poisson} \\
 Variable & estimate & $se_{eq}$ & estimate & $se$ \\
 Intercept & 3.736 & 0.026 & 3.733 & 0.020\\
 Step & -0.535 & 0.110 & -0.544 & 0.073\\
CosAnnual & 0.063 & 0.029 & 0.058 & 0.029\\
\MA\unskip, lag 1 & 0.044 & 0.017 & 0.034 & 0.016\\ 
\MA\unskip, lag 2 & -0.037 & 0.017 & -0.031 & 0.016\\ 
\end{tabular}}
\label{Table: Lockout}
\end{center}
\end{table}

The data were fit using a Poisson and the proposed semiparametric model. The coefficient estimates from the two models, along with their estimated standard errors, are reported in Table \ref{Table: Lockout}. Both models indicate that the step intervention term is highly significant and the results indicate that there has been substantial reduction in assault counts in the Kings Cross area. 

\begin{figure}
\begin{center}
\begin{tabular}{cc}
 \hskip-0.5cm \scalebox{0.35}{\includegraphics{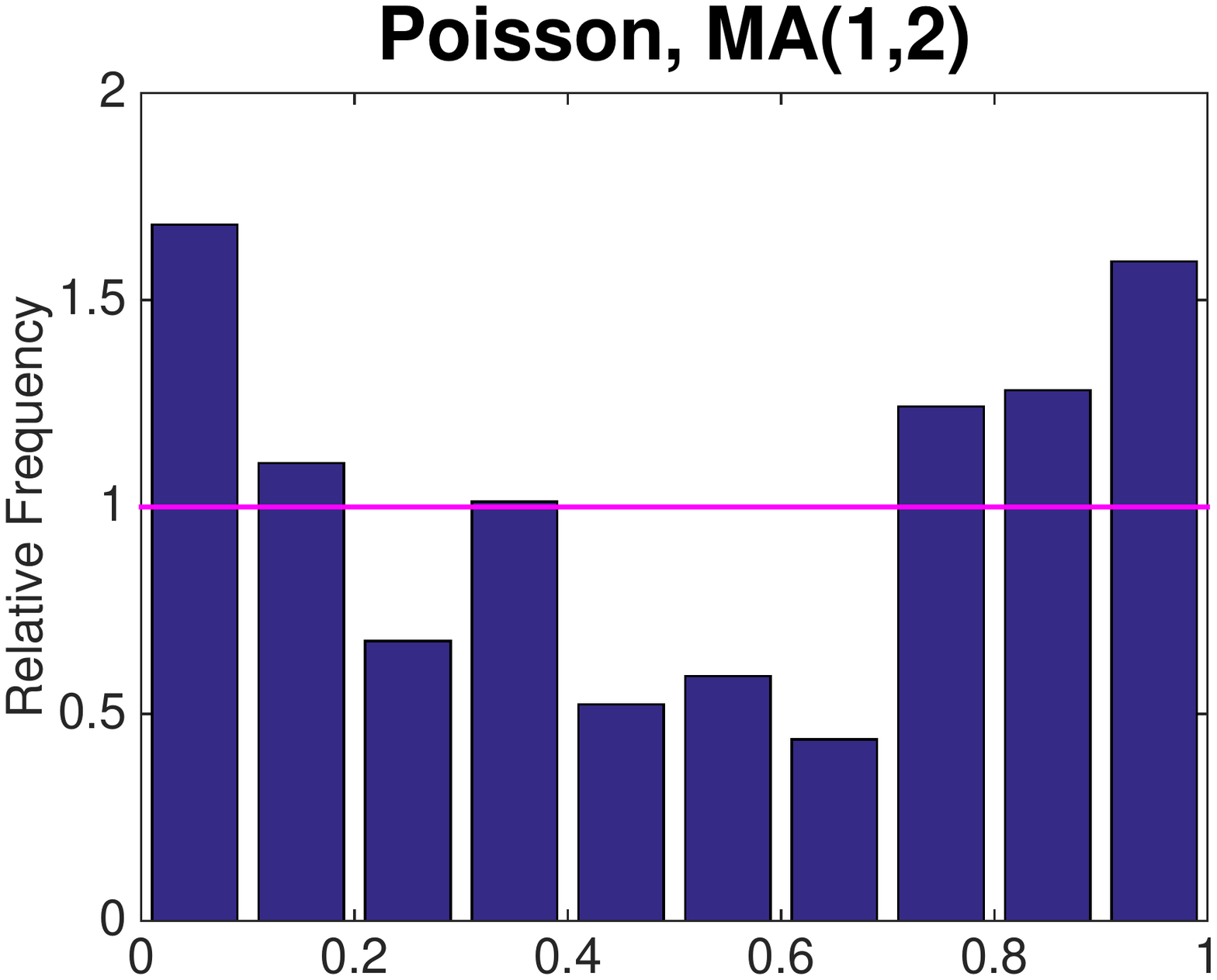}} & \hskip-0.5cm \scalebox{0.35}{\includegraphics{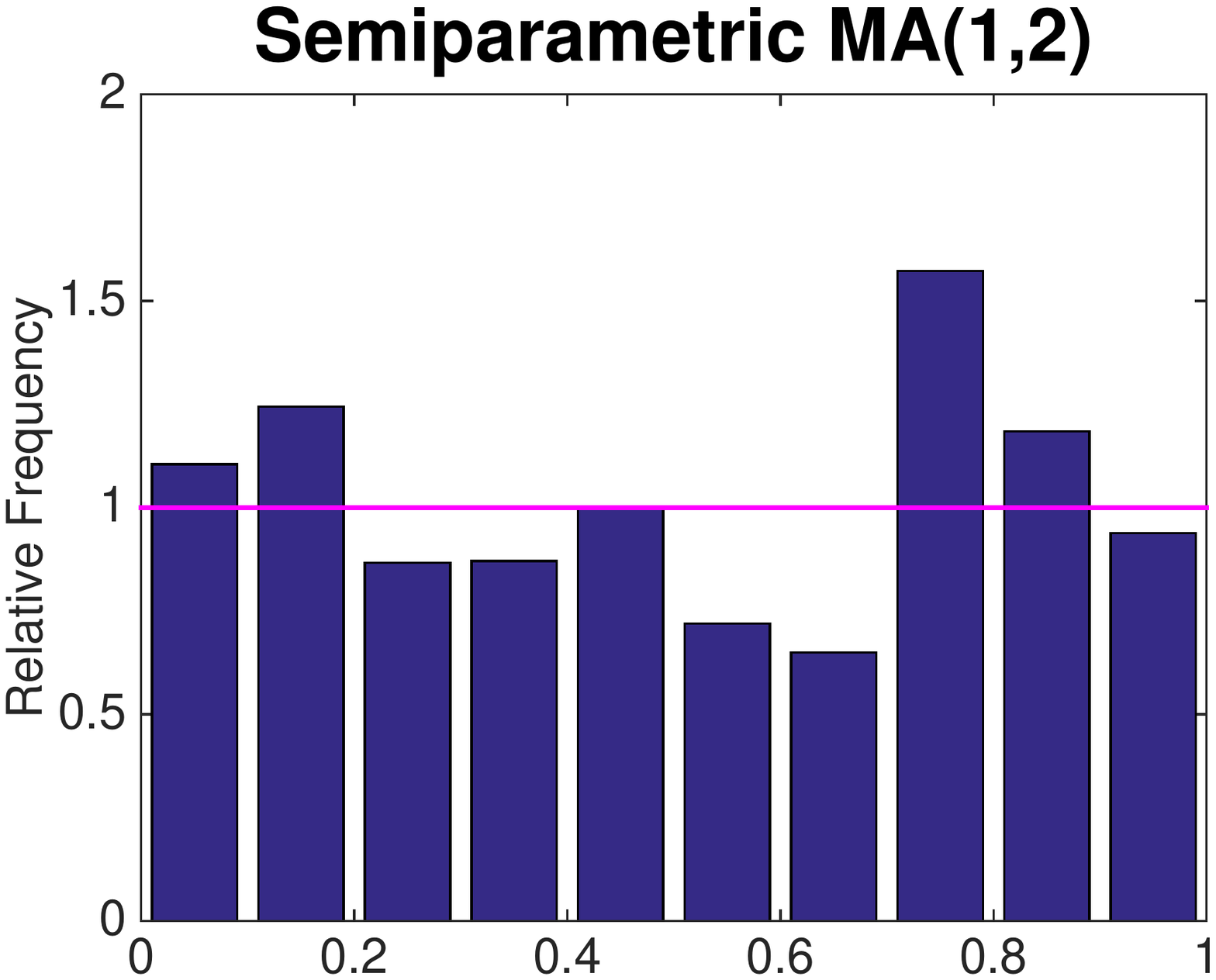}}
\end{tabular}
\end{center}
\caption{Kings Cross assault counts: Histograms of the probability integral transforms for the Poisson and semiparametric fits.} \label{figure: Lockout PIT}
\end{figure}

We can again use histograms of the \PIT to assess the Poisson distribution assumption. From the \PIT\unskip s of the two models plotted in Fig. \ref{figure: Lockout PIT} we see that there is a distinct U-shape in the plot for the Poisson model whereas the one for the semiparametric model appears closer to uniform. This suggests that the Poisson error distribution model is inadequate here, so that corresponding standard errors and subsequent inferences may well be biased. 

Interestingly, a negative-binomial \GLARMA model could not be fitted to this dataset as the Fisher Scoring algorithm failed to converge. We have yet to encounter any convergence or compatibility problems using our proposed method. Note that by leaving the underlying generating distribution unspecified, our model fitting algorithm is the same regardless of the distribution of the data. We therefore do not run into such incompatibility issues.

\section{Discussion}
\label{s:discuss}

Misspecifying the conditional response distribution in \GLARMA models can lead to biased estimation of model parameters, especially parameters governing the dependence structure. This problem of model misspecification is particularly prevalent for responses that do not have conditionally normal distributions, such as incidence counts. These types of time-series occur often in population health and epidemiology. 

The proposal in this paper is to use a semiparametric approach that maximizes over all possible exponential families of conditional distributions for the data in the \GLARMA framework. This allows for robust and flexible models for non-Gaussian time-series data, as our simulations and two data analysis examples demonstrate. The proposed approach is also a unified one, covering the class of all \GLARMA models that are based on an exponential family of conditional distributions for the responses. Moreover, the same computational algorithm can be used to fit the model to data regardless of the underlying generating distribution of the data.



\section*{Acknowledgements}

The authors also thank Professor William Dunsmuir for sharing the prerelease version of the {\tt GLARMA} software. 
\vspace*{-8pt}


\section*{Supplementary Materials}

The Online Supplement, containing additional simulation results and a proof of Theorem 1, is available by emailing the authors 
.\vspace*{-8pt}


%

\appendix


\section{\hspace{-6mm}ppendix}
\subsection{Parameter space for $F$}
As with any \textsc{glm}, the underlying distribution $F$ generating the exponential family is required to have a Laplace transform in some neighborhood of the origin, so that the cumulant generating function (\ref{eq:normconstr}) is well-defined. Note that $F$ is not identifiable since any exponentially tilted version of $F$ would generate the same exponential family. To estimate $F$ uniquely, we set $b_0 = \theta_0 = 0$ in our \texttt{MATLAB} code. 

The exponential tilt mechanism also implies that all observations $Y_t$ share a common support, $\mathcal{Y}$ -- this is another feature of \GLMs. {\color{black} Binomial responses with varying number of trials may seem to violate this, but they can be broken down into individual Bernoulli components each having common support $\{0,1\}$}, provided the number of trials at each time-point is given.  

\subsection{Regularity conditions} \label{sec:regcond}
\begin{itemize}
\item[C1.] The parameter space for $(\beta,\gamma)$ is some compact subset of $\R^{q+r}$.
\item[C2.] The response space $\mathcal{Y}$ is (contained in) some compact subset of  $\R$.
\item[C3.] The mean function $\mu(\mathcal{F}; \beta, \gamma)$ maps into the interior of $\mathcal{Y}$ for all states $\mathcal{F}$ and all parameter values $(\beta, \gamma)$ in some neighbourhood of the true value $(\beta^*, \gamma^*)$.
\end{itemize}

\begin{remark} Condition C2 is for simplicity of proof and can be weakened to tightness on the collection of allowable distributions. This then requires an additional step of approximating tight distributions by compactly-supported ones, with the compact support being increasingly widened so that the approximation error becomes arbitrarily small. 
\end{remark}

\begin{remark} Condition C3 is required because we are not restricted to the canonical link for each exponential family. For example, a linear mean model can be used with nonnegative responses, provided the mean $\mu(\mathcal{F}; \beta, \gamma)$ is never negative for any state $\mathcal{F}$, at least for models $(\beta, \gamma)$ sufficiently close to the underlying data-generating model $(\beta^*, \gamma^*)$.
\end{remark}

\label{lastpage}

\end{document}